\documentclass[english,gc, manuscript]{copernicus}
\usepackage[T1]{fontenc}
\usepackage{graphicx}
\usepackage[authoryear]{natbib}



\usepackage{babel}
\begin{document}
\title{GC Insights: Space sector careers resources in the UK need a greater
diversity of roles}
\author[1]{Martin O. Archer}
\author[1]{Cara L. Waters}
\author[1]{Shafiat Dewan}
\author[1]{Simon Foster}
\author[2]{Antonio Portas}
\affil[1]{Department of Physics, Imperial College London, London, UK}
\affil[2]{NUSTEM, Northumbria University, Newcastle, UK}
\correspondence{Martin O. Archer\\
(m.archer10@imperial.ac.uk)}
\runningtitle{Diversity of space careers resources}
\runningauthor{Archer et al.}
\maketitle
\nolinenumbers
\begin{abstract}
Educational research highlights that improved careers education is
needed to increase participation in STEM. Current UK careers resources
concerning the space sector, however, are found to perhaps not best
reflect the diversity of roles present and may in fact perpetuate
misconceptions about the usefulness of science. We, therefore, compile
a more diverse set of space-related jobs, which will be used in the
development of a new space careers resource.
\end{abstract}

\introduction{}

Educational research shows participation issues across Science Technology
Engineering and Mathematics (STEM) are not due to school students'
disinterest, but whether students see these fields and their potential
career opportunities as for ``people like me'' \citep[and references therein]{archer17}.
These perceptions form early and remain relatively stable with age,
which has led to recommendations for increased provision and quality
of careers education/engagement at both primary and secondary levels
\citep{aspires13,gatsby14,davenport20}. Careers education provision
in the UK specifically, however, is still not universal (despite mandates
being in place) and that which is provided can often be patterned
by societal inequities, unfortunately leaving some students' aspirations
``dampened'' \citep{abrahams16,archer16,moote18a,moote18b}. It
is therefore fair to say that high quality careers-related materials
are in demand by schools now more than ever.

A key problem in STEM participation is the perception that studying
science is only for those that aspire to become scientists \citep{archer17}.
This is in contrast with the wide range of careers both related to
and beyond science that further STEM education can enable. Therefore,
this ``science = scientists'' link needs to be broken by highlighting
to young people and their key influencers (e.g. teachers, parents/carers,
community leaders) the prevalence and relevance of STEM subjects to
everyday life and a diverse selection of potential career paths \citep{archer17,davenport20,archer20r4a}.

Good practice towards diversity in communications more generally may
be gleaned from the numerous efforts aimed at improving the diversity
within STEM fields, due to the under-representation of women, disabled
people, and those from ethnic-minorities or socially-disadvantaged
groups \citep{case14}. One common approach is to strive for equal
representation of minority demographics, in for instance role models,
so that those aspiring towards STEM can see ``people like me'' in
those fields which may help tackle damaging stereotypes \citep[e.g.][]{huntoon07,prinsley16,gonzalezperez20}.
In other words, equal weight should be given to all categories, irrespective
of whether they constitute a majority or minority within society.

In this paper, we investigate the representation of the space sector
within current careers resources to ascertain whether they align with
these educational recommendations.

\section{Current space careers resources}

The space sector involves a wide range of upstream (making and sending
objects to space), downstream (using these objects to deliver products/services
for exploitation), and ancillary (providing specialised support) roles,
with scientific activities spanning all three \citep{londoneconomics19,spaceskillssurvey21,knowspace21}.
In the UK alone there are 45,100 space-related roles in industry (0.14\%
of the workforce), which support 126,300 jobs across the supply chain
and generate \textsterling6.6 billion (0.30\% of the gross domestic
product). The UK space sector is currently undergoing rapid development
with many emerging opportunities (e.g. spaceports) that aim to drive
further economic benefits. However, this can only be realised if there
is a workforce trained and willing to undertake these new roles, highlighting
the need for representation of space sector roles in careers education.

\begin{figure*}
\centering{}\includegraphics{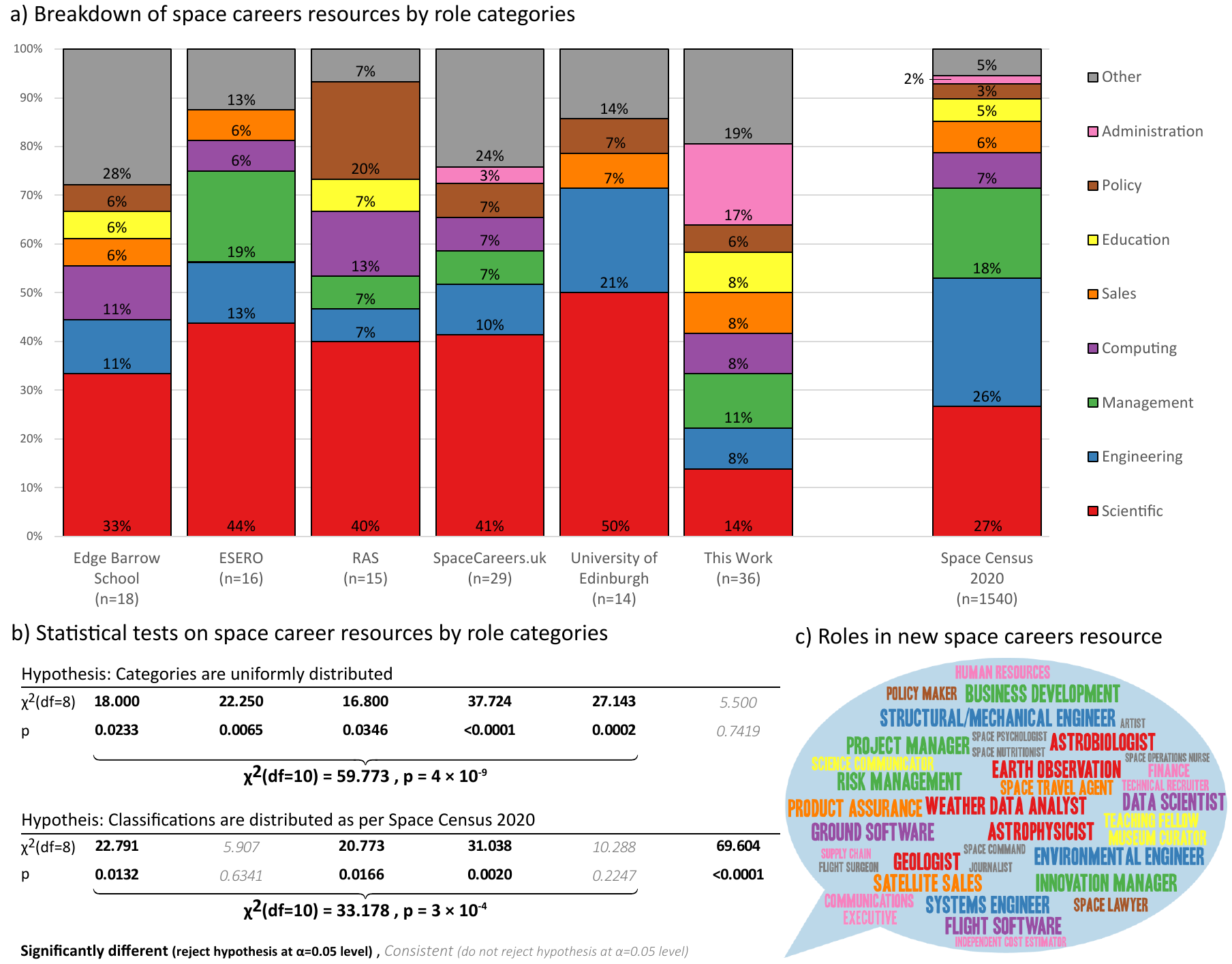}\caption{a) Breakdown of current UK space sector careers resources compared
to the UK Space Census 2020 \citep{spacecensus2020}. b) Outcomes
of chi-squared statistical tests from these distributions. c) Word
cloud of the space roles chosen for a new resource, where colours
relate to the categories in panel~a (font sizes have no meaning).\label{fig:classifications}}
\end{figure*}

The 2020 Space Census was the first national survey of the UK space
workforce \citep{spacecensus2020}, inclusive of both industry and
academia. It provides, to our knowledge, the best current classification
scheme and breakdown of the diverse roles present within the UK space
sector. These are shown to the right of Figure~\ref{fig:classifications}a.
This scheme and data are used as a benchmark for assessing current
space careers resources.

We undertook desk research to find what careers resources for young
people currently exist within the UK that aim to raise awareness and
describe a range of roles across the space sector. Our search criteria
meant that we could not include resources targeted at other countries
(e.g. USA; \citealp{bls16}), which promoted careers within specific
organisations \citep[e.g.][]{serco}, focused on only one aspect of
the sector \citep[e.g.][]{thisisengineering}, simply listed current
vacancies \citep[e.g.][]{careersinspace}, or just direct readers
to other organisations \citep[e.g.][]{uksa}. Only five sets with
at least 9 roles (i.e. 1 per space census category if evenly distributed)
were found: \citet{edgebarrowschool}, \citet{esero}, \citet{ras},
\citet{spacecareersuk}, and \citet{edinburgh16}. If other resources
within our criteria exist, they likely do not have considerable reach
or impact. The jobs featured in each of the resources were classified
using the space census scheme, performed independently by two people
finding 91\% agreement (Cohen's $\kappa=0.9$, see Appendix~\ref{sec:Statistical-methods}).
Breakdowns of each set of resources by category are shown as the first
four stacked plots in Figure~\ref{fig:classifications}a. We tested
whether the resources either had an even split of categories, which
would best reflect diversity, or were representative of the space
census. This was done through chi-squared statistical tests (see Appendix~\ref{sec:Statistical-methods}),
with the outcomes listed in Figure~\ref{fig:classifications}b below
each stacked plot.

Our results show that, to high confidence, none of the current space
careers resources have a near-even split of 11\% per category (corresponding
to 1--3 roles). Indeed, between 2--4 of the categories are missing
in each resource. Combining these tests into an overall result (see
Appendix~\ref{sec:Statistical-methods}) shows this conclusion is
highly robust. Therefore, current resources are perhaps not best representing
the diversity of space-related careers available.

Comparing the resources to the space census, we find that all of them
over-represent scientific careers. Given the low levels of young people
aspiring towards being a scientist from an early age \citep{archer17},
it appears that these resources may perpetuate misconceptions about
the usefulness of science. On the other hand, the large proportions
of ``Other'' careers across most sets means that several less traditional
career options related to space are being highlighted, which is advantageous.
The statistics indicate the Edge Barrow School, RAS, and SpaceCareers.uk
resources are highly unrepresentative of the UK space sector. We cannot
confidently claim this for the others, though they have relatively
small numbers of roles. Nonetheless, combining the results again yields
a highly significant conclusion that current space careers resources
are generally unrepresentative of the sector.

Finally, we note that these resources tend to be targeted at upper-secondary
and university students. Therefore, there appears to be a lack of
space-related careers material aimed at the ages most in need of engagement,
i.e. primary and lower-secondary students \citep{aspires13,gatsby14,davenport20}.

\section{Developing a new resource}

Given these findings, we endeavoured to create a more diverse set
of UK-based space careers for a new resource to be aimed at younger
ages. This was achieved by contacting Imperial Space Lab's industrial
partners, reading reports on the UK space sector \citep[e.g.][]{londoneconomics19,spaceskillssurvey21,knowspace21},
finding advertised vacancies, and more general online research. The
list of roles was iterated several times until it was felt the final
set of 36 careers displayed in Figure~\ref{fig:classifications}c,
greater in number than current resources, well captured the diversity
of the sector.

Our aim was that this set would have near-equal numbers in each job
category. The breakdown is shown as the fifth stacked plot in Figure~\ref{fig:classifications}a
along with results of the statistical tests (panel~b). These reveal
that our set is indeed consistent with this aim, hence better represents
the sector's diversity. Consequently, it is significantly different
from the space census, though importantly no majority category from
the census is over-represented. As with existing resources, ``Other''
careers form a significant fraction of the set thereby highlighting
less traditional paths. It is also worth noting that the high number
of roles in administration, i.e. relating to the running of a business
or organisation, was deliberate since ``business'' is by far the
most popular aspiration amongst young people \citep{aspires13}. The
application of these compiled roles into the design of a new careers
resource is beyond the scope of this paper.

\conclusions{}

Educational research has revealed improved careers education, particularly
for younger ages, may be required to improve participation in STEM.
This needs to highlight the diversity of career options STEM subjects
can enable, breaking the misconception that science is only for scientists.
Focusing on space-related careers, we have found that currently available
UK resources perhaps do not best represent the diversity of roles
present in the sector. In particular, there is an over-representation
of scientists within them, which may perpetuate stereotypes. We have,
therefore, compiled a more diverse set of space-related careers which
does not appear to suffer from these issues. These roles will form
the basis of a new space careers resource for primary and lower-secondary
students, which we hope will better align with the recommendations
from recent educational research.

\appendix

\section{Statistical methods\label{sec:Statistical-methods}}

Cohen's $\kappa$ is a measure of reliability for coding categorical
items \citep{mchugh12}. It is calculated as
\[
\kappa=\frac{p_{0}-p_{e}}{1-p_{0}}
\]
where $p_{0}$ is the proportional agreement among coders and $p_{e}$
is that expected by chance. $\kappa$ ranges between 0 (consistent
with random) and 1 (perfect agreement).

A chi-squared test compares observed frequencies $O_{i}$ within $k$
categories to those expected $E_{i}$ under some (null) hypothesis.
The statistic is given by
\[
\chi^{2}\left(\mathrm{df}=k-1\right)=\sum_{i=1}^{k}\frac{\left(O_{i}-E_{i}\right)^{2}}{E_{i}}
\]
where $\mathrm{df}$ are the degrees of freedom. Due to small numbers,
$p$-values (the probability of obtaining test results at least as
extreme as those observed) were computed via 10,000 Monte Carlo simulations
of $\chi^{2}$ for each resource's size under the hypotheses. If $p<0.05$
then the observations are considered significantly different.

$p$-values of $n$ independent tests for the same hypothesis can
be combined using \citeauthor{fisher25}'s \citeyearpar{fisher25}
method to arrive at an overall chi-squared statistic
\[
\chi^{2}\left(\mathrm{df}=2n\right)=-2\sum_{i=1}^{n}\ln\left(p_{i}\right)
\]
whose $p$-value can be calculated.

\dataavailability{Data supporting the findings are derived from listed
public domain resources.}

\authorcontribution{MOA was involved in the conceptualization, funding
acquisition, supervision, formal analysis, visualization, and writing
of this work. CLW and SD designed the methodology, performed the investigation,
and undertook data curation. SF contributed to project administration
and supervision. AP provided resources and assisted with validation.}

\competinginterests{The authors declare that they have no conflict
of interest.}
\begin{acknowledgements}
We thank the Jonathan Eastwood for his support in this project and
research. This work has been made possible by Imperial College London's
Undergraduate Research Opportunity Programme, through funding from
the Department of Physics and Space Lab. M.O. Archer holds a UKRI
(STFC / EPSRC) Stephen Hawking Fellowship EP/T01735X/1. We are also
thankful for funding from STFC (ST/W00545X/1).
\end{acknowledgements}
\bibliographystyle{copernicus}
\bibliography{spacecareersinsight}

\end{document}